# Pressure-induced Superconductivity in the Three-component Fermion Topological Semimetal Molybdenum Phosphide

Zhenhua Chi[1], Xuliang Chen[2], Chao An[2,3], Liuxiang Yang[4], Jinggeng Zhao[5,6], Zili Feng[7,8], Yonghui Zhou[2], Ying Zhou[2,3], Chuanchuan Gu[2], Bowen Zhang[2,3], Yifang Yuan[2,9], Curtis Kenney-Benson[10], Wenge Yang[11], Gang Wu[12], Xiangang Wan[13,14], Youguo Shi[7,¡], Xiaoping Yang[2,$], Zhaorong Yang[2,14,§]

[1]Key Laboratory of Materials Physics, Institute of Solid State Physics, Chinese Academy of Sciences, Hefei 230031, China

[2]Anhui Province Key Laboratory of Condensed Matter Physics at Extreme Conditions, High Magnetic Field Laboratory, Chinese Academy of Sciences, Hefei 230031, China

[3]University of Science and Technology of China, Hefei 230026, China

[4]Center for High Pressure Science and Technology Advanced Research (HPSTAR), Beijing 100094, China

[5]Department of Physics, Harbin Institute of Technology, Harbin 150080, China

[6]Natural Science Research Center, Academy of Fundamental and Interdisciplinary Sciences, Harbin Institute of Technology, Harbin 150080, China

[7]Beijing National Laboratory for Condensed Matter Physics and Institute of Physics, Chinese Academy of Sciences, Beijing 100190, China

[8]School of Physical Sciences, University of Chinese Academy of Sciences, Beijing 100190, China

[9]Department of Physics and Engineering, Zhengzhou University, Zhengzhou 450052, China

[10]HPCAT, Geophysical Laboratory, Carnegie Institution of Washington, Argonne, Illinois 60439, USA

[11]Center for High Pressure Science and Technology Advanced Research (HPSTAR), Shanghai 201203, China

[12]Institute of High Performance Computing, 1 Fusionopolis Way, 16-16 Connexis, Singapore 138632, Singapore

[13]National Laboratory of Solid State Microstructures, College of Physics, Nanjing University, Nanjing 210093, China

[14]Collaborative Innovation Center of Advanced Microstructures, Nanjing University, Nanjing 210093, China

----------------------------------------------------------------------------------

E-mails: §zryang@issp.ac.cn; $xpyang@hmfl.ac.cn; ¡ygshi@iphy.ac.cn

**Abstract**

Topological semimetal, a novel state of quantum matter hosting exotic emergent quantum phenomena dictated by the non-trivial band topology, has emerged as a new frontier in condensed-matter physics. Very recently, a coexistence of triply degenerate points of band crossing and Weyl points near the Fermi level was theoretically predicted and immediately experimentally verified in single crystalline molybdenum phosphide (MoP). Here we show in this material the high-pressure electronic transport and synchrotron X-ray diffraction (XRD) measurements, combined with density functional theory (DFT) calculations. We report the emergence of pressure-induced superconductivity in MoP with a critical temperature $T_c$ of about 2 K at 27.6 GPa, rising to 3.7 K at the highest pressure of 95.0 GPa studied. No structural phase transitions is detected up to 60.6 GPa from the XRD. Meanwhile, the Weyl points and triply degenerate points topologically protected by the crystal symmetry are retained at high pressure as revealed by our DFT calculations. The coexistence of three-component fermion and superconductivity in heavily pressurized MoP offers an excellent platform to study the interplay between topological phase of matter and superconductivity.

The low-energy, long-wavelength quasiparticle excitation in the vicinity of Fermi level as counterpart of relativistic Dirac, Weyl and Majorana fermion in high-energy physics has been successfully demonstrated in topological states of quantum matter with non-trivial band topology such as topological insulator (TI), Dirac semimetal (DS), Weyl semimetal (WS), and topological superconductor (TSC) over the last decade [1-14]. In stark contrast with gapped bulk state of both TI and TSC, the defining hallmark of DS and WS is the presence of linear crossing points of conduction bands and valence bands near the Fermi level, where two (WS) or four bands (DS) are exactly degenerate at certain momentum-energy value in the first Brillouin zone, giving rise to a gapless bulk state. Novel macroscopic quantum phenomena they exhibited such as ultrahigh mobility of charge carrier, extremely large linear magnetoresistance, quantum anomalous Hall effect and chiral anomaly are not only of fundamental interest but also hold potential applications. The recent breakthrough in predicting and identifying unconventional "new fermion" beyond conventional Dirac and Weyl fermion has sparked new research interest in the field of topological semimetal [15-20]. Among the symmetries of 230 crystal space groups in condensed-matter system, the threefold-, sixfold- and eightfold degenerate quasiparticle excitation emanating from the multiply degenerate points of band crossing near the Fermi level can be protected in a crystal lattice either by symmorphic rotation combined with mirror symmetries or by non-symmorphic symmetries. The novel three-component, six-component and eight-component fermions have distinct properties from conventional Dirac and Weyl one, including unique surface states and transport properties. One such material hosting the triply degenerate fermion which has been experimentally identified via angle-resolved photoemission spectroscopy (ARPES) is WC-structured MoP with the symmorphic space group $P6m2$ (number 187) [19]. The rotational symmetry $C_{3z}$ and the mirror symmetries $M_y$ and $M_z$ are crucial to topologically protect the triply degenerate points in MoP.

Along with the quest of exotic "new fermion", the materialization of topological superconductivity which is topologically distinct from conventional Bose-Einstein condensates of Cooper pairs has also constituted an active research theme. As a manifestation of the topological character, TSC promises solid-state realization of itinerant massless Majorana fermion which may operate as topological qubit for fault-tolerant quantum computation. However, the natural occurrence of TSC is extremely scarce with the unique promising candidate being $Sr_2RuO_4$ for its spin-triplet pairing [21], despite that a wide range of candidate materials have been proposed [22-25]. Turning the topological phase of matter into superconducting state by either doping or pressurizing is believed to be a practical avenue to access the TSC [26-43], however, the topological non-trivial nature of the induced superconductivity is under heavy debate yet. Pressure, modifying the crystal structure and electronic band structure in a systematic fashion, is a clean and efficient thermodynamic parameter in tuning the physical properties of topological quantum matters [36-43, 44]. Here, via high-pressure electronic transport and synchrotron X-ray diffraction measurements with the aid of theoretical calculations, we present unambiguous evidences of pressure-induced superconductivity in the newly identified three-component fermion topological semimetal MoP, offering a new paradigm with coexistence of superconductivity and topological fingerprint.

## Results

**Crystal structure characterization at ambient pressure.** The single crystal sample was synthesized by Ge-flux method as described in ref [19] and also has been studied by ARPES. Single crystal X-Ray diffraction at ambient pressure was measured on Bruker D8 Venture by using Mo $K\alpha_1$ radiation ($\lambda = 0.71073$ Å) at 300 K and the data was refined using the SHELXL-97 program. The space group is *P-6m2* (No. 187) and the lattice constants are $a = b = 3.2259(3)$ Å, $c = 3.2050(3)$ Å, $\alpha = \beta = 90°$ and $\gamma = 120°$. Figure 1 shows the diffraction peaks from (001) plane which are well-defined and indexed, indicating high-quality crystallization and perfect surface of (001) plane.

**Pressure-induced superconductivity.** The temperature-dependent resistance at various pressures in two runs are shown in Fig. 2. As shown in Fig. 2a for run 1, at pressures below 23 GPa, the sample displays a conducting behavior characteristic of a normal semimetal down to the lowest temperature measured, indicated by the positive sign of d$R$/d$T$. At 30 GPa, although the overall conduction is metallic, a slight upturn in resistance occurs upon cooling below a characteristic temperature $T^*$ of 2.5 K, as seen in Fig.2b. This slight upturn in resistance persists up to 41.1 GPa. As shown in Fig. S1, a magnetic field of 0.2 T can almost smear out this resistance upturn, signaling the appearance of superconductivity with a transition temperature beyond the lower limit of our cryostat. In the high-pressure electronic transport measurement, the resistance sometimes displays a slight upturn preceding the drop in the vicinity of the superconducting transition [37-39, 43, 45, 46]. Such an abnormal effect has also been observed in mesoscopic superconducting disk, and attributed to the charge imbalance effects at a superconductor/normal metal junction [47]. Upon further increasing the pressure, superconductivity characterized by a slight drop in the resistance emerges with a transition temperature $T_c$ of 2.5 K at 46.7 GPa, as shown in Fig. 2b. The $T_c$ is defined at the point where the resistance starts to deviate from the nearly temperature-independent constant in the normal state. The magnitude of drop in the resistance is progressively enhanced with pressure increasing up to 61.5 GPa. Simultaneously, the $T_c$ is lifted up to 3 K. To unambiguously confirm the superconducting nature of this resistive transition, we repeated the measurement at pressures up to 100 GPa aiming at obtaining zero resistance. As shown in Fig. 2c for run 2, at pressures below 21.5 GPa, the sample behaves as a normal semimetal without marked signature of superconductivity down to the lowest temperature measured. Upon further increasing the pressure to 27.6 GPa, superconductivity characterized by a slight drop in the resistance sets in with a transition temperature $T_c$ of 2 K, as shown in Fig. 2d. With pressure increasing, the $T_c$ rises monotonically up to 3.7 K at 95 GPa. At the highest pressure of 95 GPa achievable in this measurement, zero resistance is explicitly observed, substantiating the occurrence of superconductivity.

**Determination of the upper critical field.** To further substantiate that the precipitous drop in the temperature-dependent resistance is of superconducting origin, the effect of external magnetic field on this resistive transition was investigated at 95 GPa under external magnetic field up to 1 T, as shown in Fig. 3a. The drop in the resistance is significantly suppressed with increasing the magnitude of external magnetic field. As shown in Fig. 3b, the temperature dependence of upper critical field $H_{c2}$ defined as 90％of the normal state resistance is well fitted by a modified Ginzburg-Landau equation $H_{c2}(T)=H_{c2}(0)*(1-T/T_c)^{(1+\alpha)}$ with the zero-temperature upper critical field $H_{c2}(0)$ of 3.75 T, which is far below the BCS Pauli limit of $H_p$ (T)=1.84$T_c$ (K)=6.8 T with $T_c$=3.7 K, indicating the absence of Pauli paring. The upward curvature of $H_{c2}(T)$ close to $T_c$ hints at a multiband superconducting pairing state [48-50].

**Pressure-temperature phase diagram.** The pressure-temperature phase diagram of MoP based on the temperature-dependent resistance measurement is summarized in Fig. 4. At pressures below 27.6 GPa, MoP behaves as a normal topological semimetal without marked signature of superconductivity down to the lowest temperature measured. At 27.6 GPa, superconductivity indicated by a slight drop in the resistance emerges at around 2 K. Upon further compression, the $T_c$ increases monotonically up to 3.7 K at 95 GPa. In the superconducting region, no discontinuity is observed in the pressure dependence of resistance at 5 K, as shown in Fig. S2. In contrast, the emergence of pressure-induced superconductivity in other topological electronic materials is normally accompanied by a structural phase transition which inevitably leads to discontinuity either in the *P-T* phase diagram or in the changes of normal state resistance. The appearance of pressure-induced superconductivity in MoP is irrelevant to the structural phase transition, as confirmed by the high-pressure synchrotron X-ray diffraction study.

**Crystal structure at high pressure.** To track the structure evolution of MoP at high pressure, high-pressure synchrotron X-ray diffraction (HPXRD) was conducted at room temperature. The HPXRD patterns and results of Rietveld refinement are shown in Fig. 5. As shown in Fig. 5a, the WC-type crystal structure of MoP is stable against pressure up to 60 GPa, indicating that the origin of pressure-induced superconductivity is irrelevant to a structural phase transition. As seen in Fig. 5b, the lattice constant of *a* and *c*, volume of unit cell and Mo-P bond length shrinks monotonically without any anomaly upon pressurization. However, there undergoes a discontinuity in the pressure dependence of both the axial ratio *c/a* and the Mo-P-Mo bond angle at 25.3 GPa where the superconductivity sets in, implying a subtle correlation between superconductivity and lattice degree of freedom in MoP. Fitting the experimental *V-P* data by the Birch-Murnaghan equation of state and assuming the first-order pressure derivative $B_0'$ to be 4 for the two pressure regions below and above 25.3 GPa, the bulk modulus at ambient pressure ($B_0$) is determined to be 213(1) GPa and 238(2) GPa with the fitted volume at ambient pressure ($V_0$) of 28.852(8) Å$^3$ and 28.59(3) Å$^3$, respectively. The refined lattice constants at various pressure are shown in the Supplementary Table.

**Density functional theory (DFT) calculation of electronic band structure at high pressure.** To gain a comprehensive understanding of the pressure–induced superconductivity in MoP, we performed DFT calculations of the electronic band structures at high pressure. The results for band structure at 0 GPa shown in Fig. S3 agree well with the previous study [19], indicating that MoP is a topological semimetal with coexistence of Weyl fermion and massless triply degenerate fermion. We applied the crystal structure prediction techniques using USPEX to search for the possible high pressure equilibrium configurations. Our simulations confirm that no structural phase transition occurs around the critical pressure where superconducting transition is observed experimentally.

Our band structure calculations with spin–orbit coupling (SOC) in Figs. 6**b**–6**d** reveal that the coexisting Weyl points (WPs) and triply degenerate points (TPs) survive at high pressure and all four bands crossing the Fermi level are topologically nontrivial ones contributing to the formation of Weyl fermions and/or triply degenerate fermions. The WPs and TPs are topologically protected by the crystal symmetry which is determined to be robust at high pressure via the HPXRD study. The Fermi surfaces (FSs) of the four topological bands are plotted in Fig. 6e. Except for the two hole pockets around $\Gamma$ and $K$ points, MoP has the strong three–dimensional electron FSs. It is noted that the pressure qualitatively changes the energy position of band 6 at $K$ point from valence band to conduction band in the band structures without SOC, as shown in Fig. 6a. Here, the hybrid functional calculation is used to eliminate the possible overestimation of band inversion in GGA, and their comparison indicates that band inversions around $K$ and the pressure–dependent qualitative change of the band 6 in energy at $K$ point occur in both cases. Furthermore, the pressure–dependent evolution of band 6 without SOC and bands 11-12 with SOC in energy at $K$ point (Fig. S4) undergoes a discontinuity between 25 GPa and 30 GPa, where the subtle modification of Fermi surface and onset of superconductivity are indicated experimentally, pointing to a correlation among them.

## Discussion

As is well-established, the emergence of pressure-induced superconductivity in topological insulator $Bi_2Se_3$, Dirac semimetal $Cd_3As_2$ and Weyl semimetal TaP [43, 51, 52] is accompanied by a structural phase transition which inevitably breaks the crystal symmetry protecting the topological state. However, the pressure-induced superconductivity in MoP emerges without concomitant structural phase transition, implying the coexistence of superconductivity with topological feature in MoP at high pressure.

Theoretically, a time–reversal–invariant topological superconductor requires odd–parity symmetry and the Fermi surface enclosing an odd number of time-reversal invariant momenta (TRIM) [29]. The states at Fermi surface are constituted by the hybridization of $4d$ orbital of Mo and $5p$ orbital of P. Because these bands are spatially extended, the electronic correlation is quite small due to the strong screening effect. As we can see in Fig. 6a, the hybrid functional calculation with correlation correct does not induce significant change of the Fermi surfaces compared to GGA calculations. Consequently, one can expect that the superconductivity discovered in this work is mainly dictated by the electron–phonon interaction. However, the phonon-mediated pairing can possess odd–parity symmetry, if the electron–phonon interaction has singular behavior at long wavelengths [53]. MoP has four TRIM at $\Gamma$, $A$, $M$, $L$ in the first Brillouin zone. As shown in Fig. 6e, both Fermi surfaces of band 11 and 12 enclose one TRIM at $\Gamma$ point. As a result, the occurrence of topological superconductivity at pressure above 30 GPa in MoP is highly favored.

Our findings represent the first experimental paradigm of pressure-induced superconductivity in the newly discovered three-component fermion system, providing an excellent platform for exploring the exotic emergent quantum phenomena arising from interplay between topological properties and superconductivity.

## Methods

**High pressure electronic transport measurements.** The measurements were carried out in a screw-pressure-type diamond anvil cell (DAC) made of non-magnetic Cu-Be alloy. The tungsten gasket with initial thickness of 250 μm was preindented by a pair of diamond anvil with culet size of 200 μm to a pressure of 20 GPa. Afterwards, a hole with diameter of 200 μm was drilled by laser ablation to remove the metal part around the culet completely. The mixture of epoxy and cubic-BN powder was loaded and compressed to a pressure of 35 GPa for insulation. Pt foil with thickness of 5 μm was used for electrical probes. A flake sample was forced to contact with the four Pt probes in a van der Pauw configuration (inset of Fig. 2**d**). Low temperature measurements were carried out in a cryostat with a base temperature of 1.7 K (JANIS Research Company, Inc.). The pressure was determined by ruby fluorescence below 80 GPa [54] and by diamond Raman peak above 80 GPa at room temperature, before and after each cooling.

**High pressure synchrotron X-ray diffraction**. The experiment was performed at Sector 16 BMD, HPCAT at Advanced Photon Source, Argonne National Laboratory. A symmetric diamond anvil cell was used. Pressure was generated by a pair of diamond anvils with culet size of 300 μm. Rhenium gasket was preindented to about 40 μm in thickness and a hole with a diameter of 100 μm was drilled to serve as sample chamber. Ruby ball as pressure marker were loaded together with MoP powder ground from as-grown single crystal into the sample chamber. Neon was used as pressure-transmitting medium. A monochromatic X-ray beam with incident wavelength of 0.3100 Å was used. The diffraction patterns were collected with a MAR 345 image plate detector. The two-dimensional image plate patterns were integrated into one-dimensional intensity versus

2θ data using the Fit2D software package [55]. Refinement of the X-ray diffraction patterns was performed using the GSAS+EXPGUI software packages [56, 57].

**Density functional calculations.** The high–pressure behaviour were explored by merging the evolutionary algorithm and *ab initio* total–energy calculations, as implemented in the USPEX code [58, 59]. Enthalpy and electronic structure calculation were carried out by using the Vienna *ab initio* Simulation Package (VASP) [60, 61] within the framework of generalized gradient approximation (GGA) (Perdew-Burke-Ernzerhof exchange correlation functional) [62]. The ion–electron interaction was modeled by the projector augmented wave (PAW) method [63, 64] with a uniform energy cutoff of 330 eV. The possible underestimation of band gap within GGA is checked by a nonlocal Heyd–Scuseria–Ernzerhof (HSE06) hybrid functional calculation [65]. Spacing between k points was 0.02 Å$^{-1}$. The geometry structures were optimized by employing the conjugate gradient technique, and in the final geometry, no force on the atoms exceeded 0.001 eV/Å. Spin–orbit coupling for all elements was taken into account by a second–variation method. The projected Wannier functions for the Mo–4d and P–5p are generated by using Wannier90 [66, 67]. A tight–binding model based on these has been established to calculate the Fermi surfaces and the position of Weyl nodal points in the Brillouin zone.

## Acknowledgments

This work is supported by the National Key Research and Development Program of China (Grants No. 2016YFA0401804, No. 2017YFA0302901, No. 2017YFA0403600), the National Natural Science Foundation of China (Grants No. U1632275, No. 11574323, No. 51372249, No. 11674328, No. 11374307, No. 11404343, No. 11774399, No. 11474330, No. 11674325, No. 11644001, No. U1632162), and the Natural Science Foundation of Heilongjiang Province (Grant No. A2017004). The XRD was performed at HPCAT (Sector 16), Advanced Photon Source, Argonne National Laboratory. HPCAT operations are supported by DOE-NNSA under Award No. DE-NA0001974 and DOE-BES under Award No. DE-FG02-99ER45775, with partial instrumentation funding by NSF. The Advanced Photon Source is a U.S. Department of Energy (DOE) Office of Science User Facility operated for the DOE Office of Science by Argonne National Laboratory under Contract No. DE-AC02-06CH11357.

## Additional information

**Competing financial interests**: The authors declare no competing financial interests.

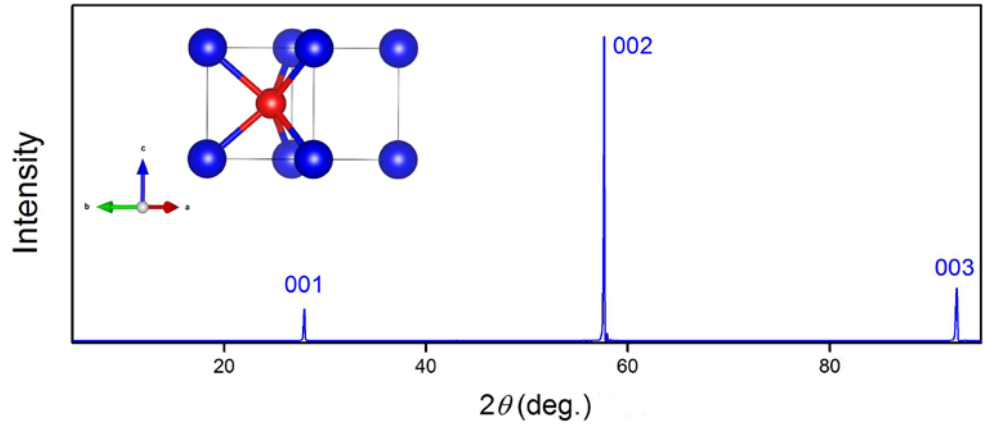

**Figure 1| Structural characterization of as-grown MoP single crystal at ambient pressure.** The XRD pattern of (00*l*) plane of as-grown MoP single crystal. Inset illustrates the hexagonal WC-type crystal structure of MoP, in which the red ball represents the P atom and the blue balls represent the Mo atom, respectively.

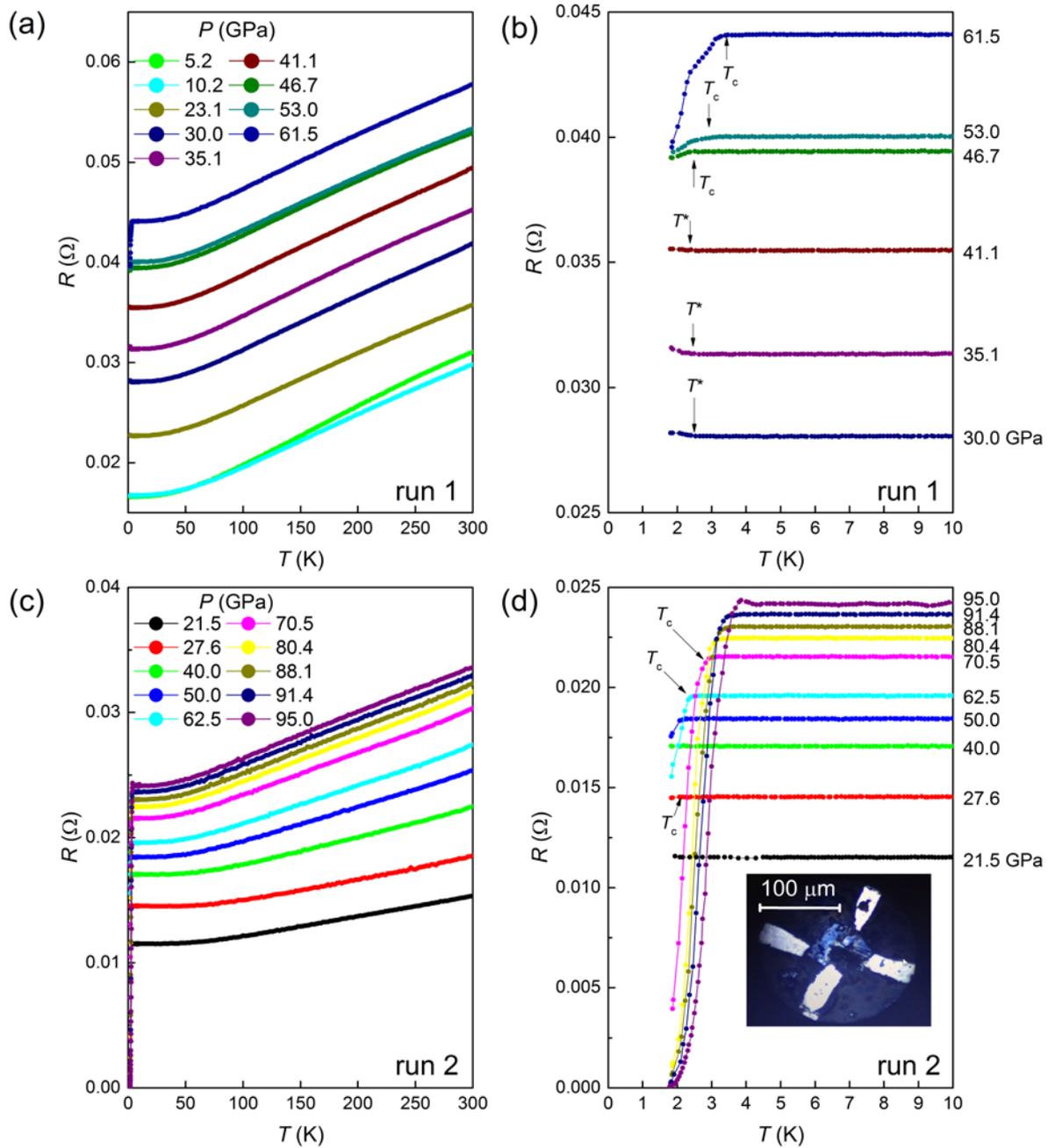

**Figure 2|Temperature-dependent resistance of MoP at various pressures.** *R-T* curves in the whole temperature range for run 1 **a** and run 2 **c**. An enlarged view of the low temperature range for run 1 **b** and run 2 **d**. In Figs. 2b and 2d, the characteristic temperature $T^*$ and superconducting transition temperature $T_c$ are indicated by arrow. Inset of Fig. 2d displays the configuration of the electrodes inside the diamond anvil cell.

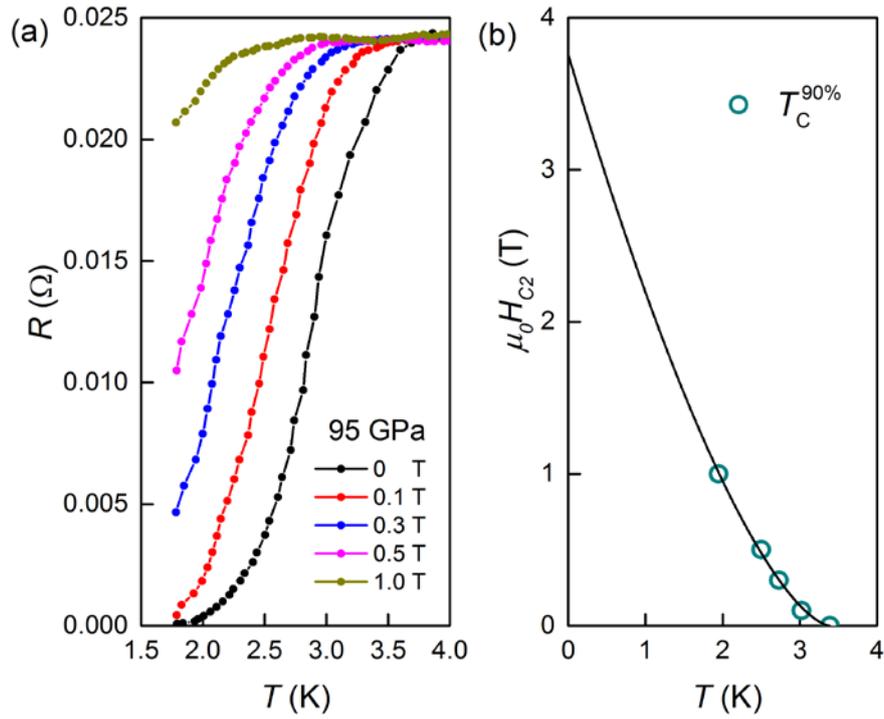

**Figure 3|Determination of the upper critical field of MoP at 95 GPa. a,** *R-T* curve under applied magnetic fields up to 1 T. **b,** $H_{c2}$-*T* phase diagram. The solid line represents the best fitting by the modified G-L equation $H_{c2}(T)=H_{c2}(0)(1-T/T_c)^{(1+\alpha)}$.

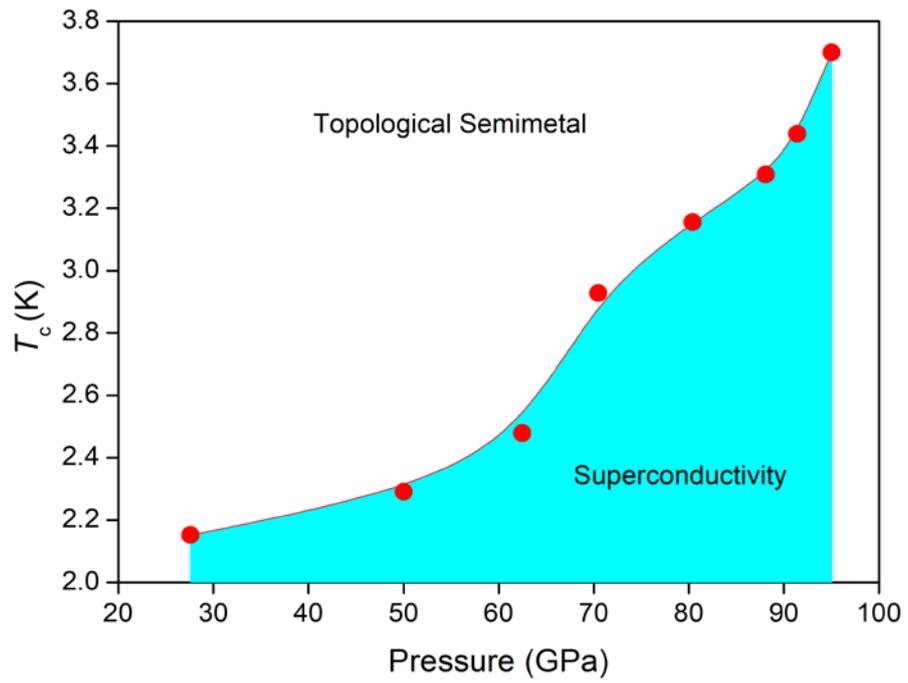

**Figure 4|Pressure-temperature phase diagram of MoP.**

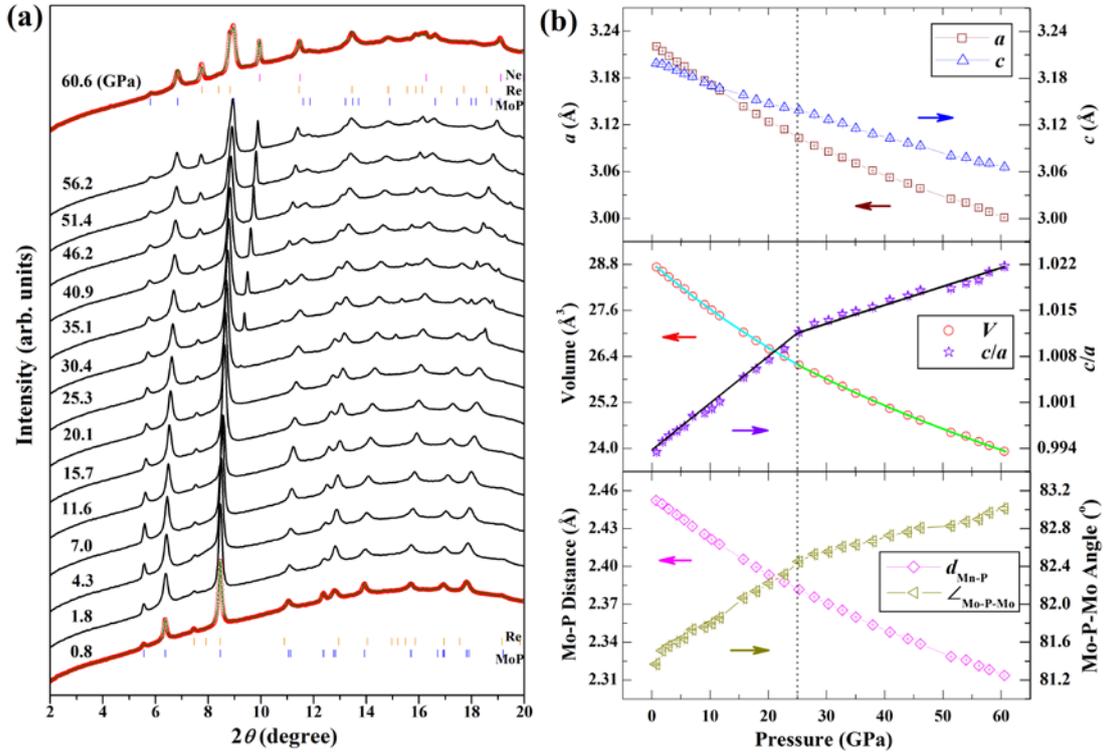

**Figure 5|High-pressure synchrotron X-ray diffraction analysis of MoP. a,** X-ray diffraction patterns of MoP up to 60.6 GPa at room temperature ($\lambda$=0.31 Å). Experimental (open circle) and fitted (line) XRD patterns at 0.8 GPa and 60.6 GPa are shown, with the $R_{wp}$ factor of 0.62% and 0.86%, respectively, in which the vertical lines denote the theoretical positions of the Bragg peaks. **b,** Pressure dependence of lattice constant (*a* and *c*), volume (*V*), axial ratio (*c/a*), Mo−P bond length, and Mo−P−Mo bond angle of MoP. The cryn and green bold lines in the *V-P* curve represent the fitting to experimental data by the Birch-Murnaghan equation of state.

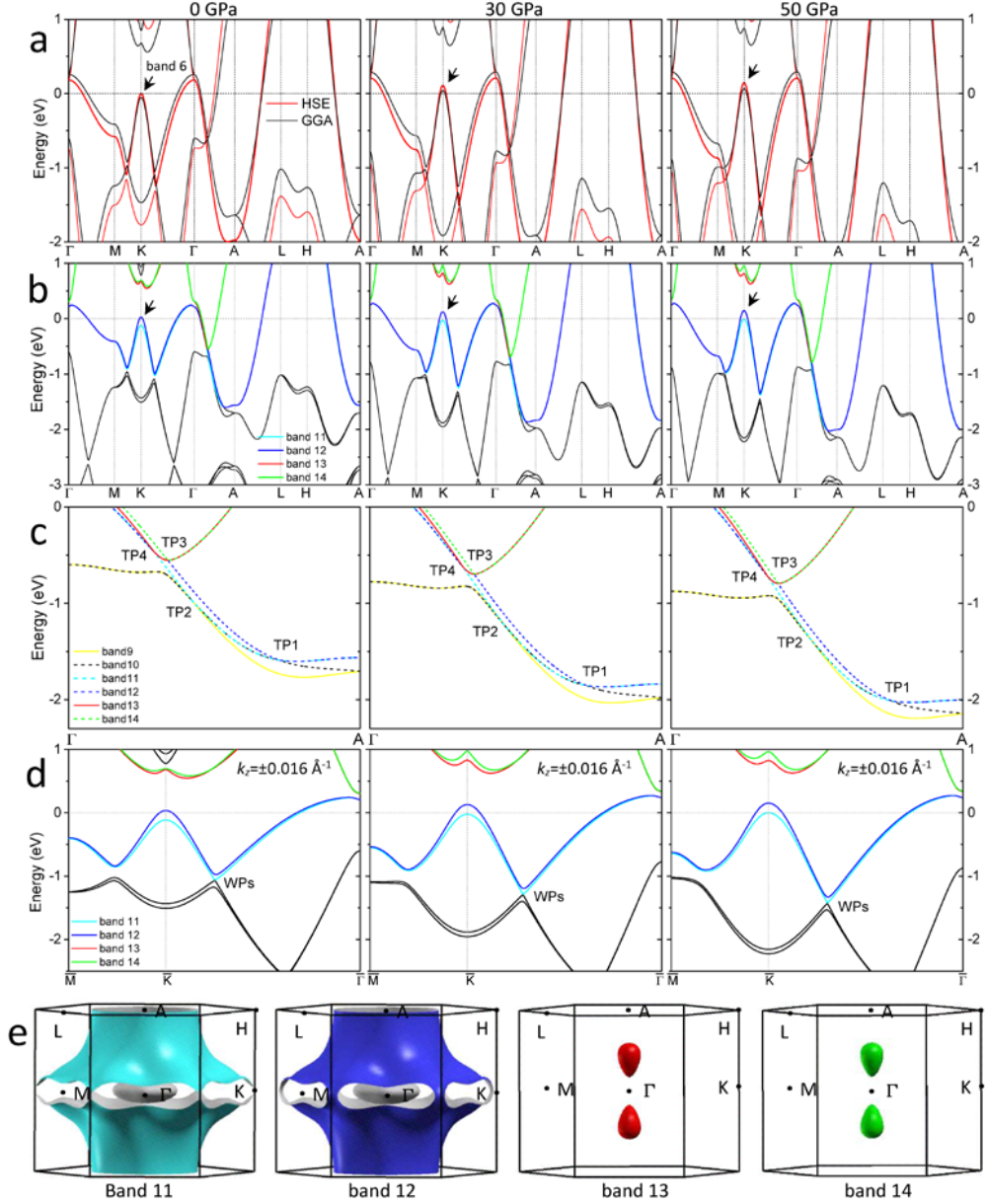

**Figure 6|The electronic band structure of MoP at different pressure. a.** The band structures calculated without SOC within GGA (black lines) and hybrid functional HSE06 (red lines). **b**. The GGA band structures with SOC. **c**. The GGA band structures with SOC around triply degenerate nodes along $\Gamma$-$A$ line. **d**. The GGA band structures with SOC around Weyl nodes at $k_z= \pm 0.016$ Å$^{-1}$ planes. The Fermi level is set at zero in all band structures. **e**. The Fermi surfaces of the four bands (crossing the Fermi level) with SOC for MoP at 30 GPa. The first two bands enclose one TRIM $\Gamma$ point.

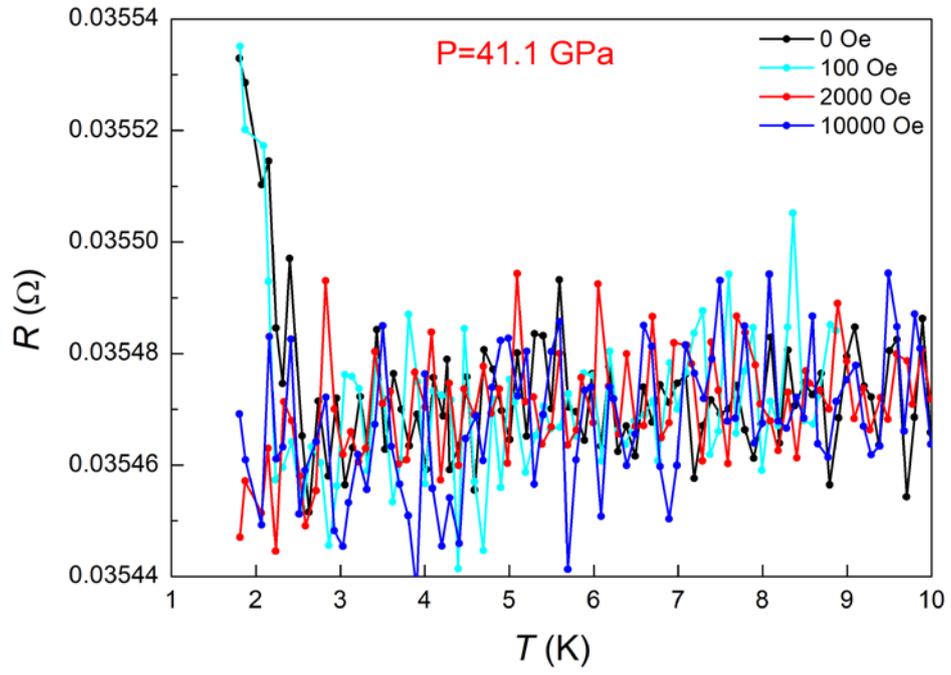

**FIG. S1. (Color online) The suppression of resistive transition with application of external magnetic field at 41.1 GPa in run 1.**

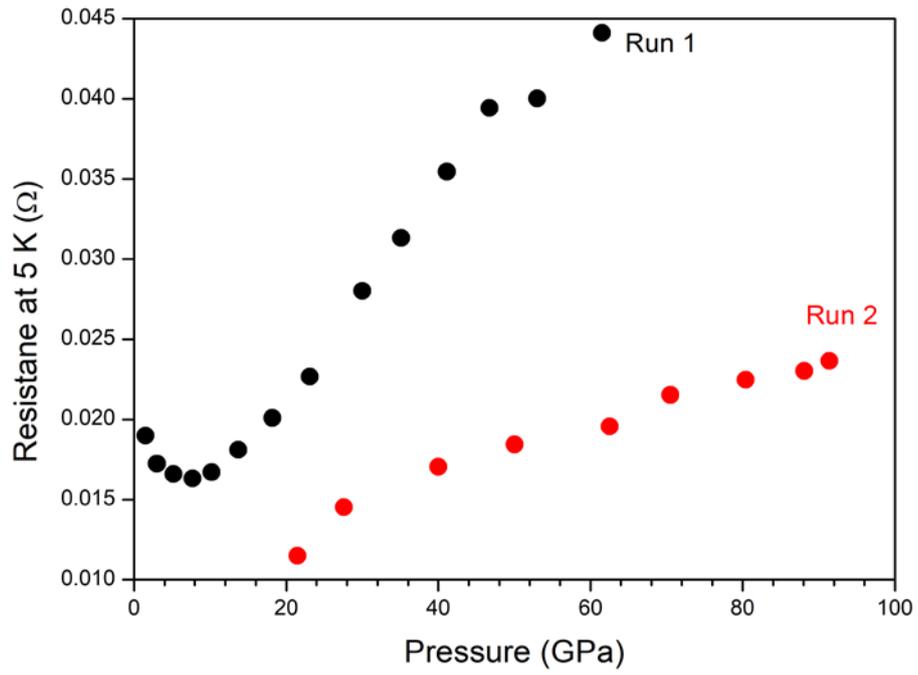

**FIG. S2. (Color online) Resistance at 5 K as a function of pressure in two runs.**

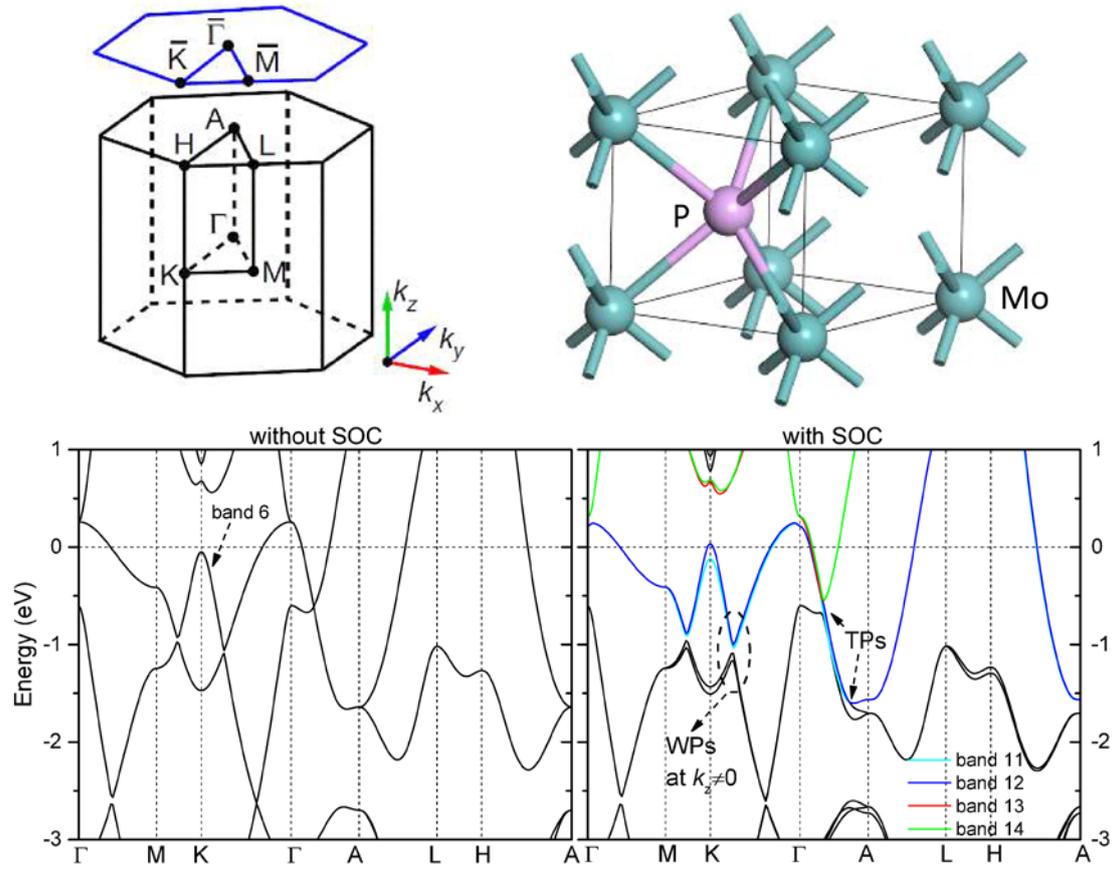

**FIG. S3. (Color online)** Upper panel: Three-dimensional crystal structure of MoP and its bulk Brillouin zone (black) and the projected (001) surface Brillouin zone (blue), with high-symmetry points indicated. Lower panel: The band structures calculated without and with SOC within GGA. The Fermi level is set at zero.

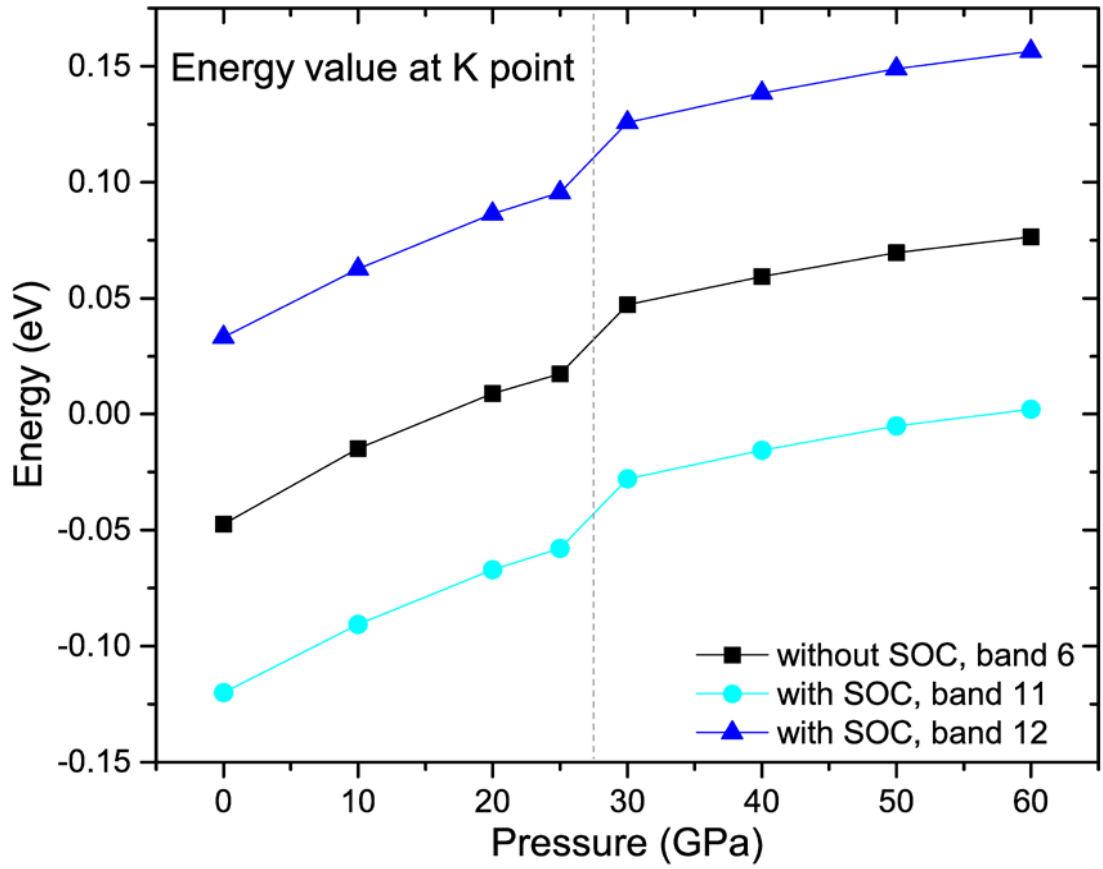

**FIG. S4.** (**Color online**) The pressure-dependent evolution of band 6 without SOC and bands 11-12 with SOC in energy at *K* point.

**Table I**. Refined lattice constants of MoP at various pressures.

| P (GPa) | a (Å) | c (Å) | $R_p$ | $R_{wp}$ |
|---|---|---|---|---|
| 0.8 | 3.2203(2) | 3.1993(4) | 0.48% | 0.62% |
| 1.8 | 3.2143(3) | 3.1983(5) | 0.74% | 1.11% |
| 2.9 | 3.2078(3) | 3.1947(6) | 0.80% | 1.26% |
| 4.3 | 3.2008(4) | 3.1901(6) | 1.00% | 1.64% |
| 5.6 | 3.1945(2) | 3.1862(3) | 0.93% | 1.23% |
| 7.0 | 3.1855(2) | 3.1822(3) | 0.73% | 0.92% |
| 9.1 | 3.1768(2) | 3.1750(3) | 0.66% | 0.82% |
| 10.3 | 3.1705(2) | 3.1708(4) | 0.63% | 0.83% |
| 11.6 | 3.1639(2) | 3.1676(4) | 0.62% | 0.78% |
| 15.7 | 3.1434(2) | 3.1586(4) | 0.53% | 0.68% |
| 17.9 | 3.1336(2) | 3.1528(4) | 0.49% | 0.57% |
| 20.1 | 3.1237(2) | 3.1473(4) | 0.49% | 0.56% |
| 22.8 | 3.1143(2) | 3.1430(4) | 0.62% | 0.73% |
| 25.3 | 3.1030(2) | 3.1393(5) | 0.60% | 0.69% |
| 27.9 | 3.0934(3) | 3.1337(5) | 0.61% | 0.69% |
| 30.4 | 3.0856(3) | 3.1273(5) | 0.64% | 0.75% |
| 32.8 | 3.0781(3) | 3.1226(5) | 0.68% | 0.83% |
| 35.1 | 3.0704(3) | 3.1160(5) | 0.68% | 0.72% |
| 38.0 | 3.0614(3) | 3.1089(6) | 0.65% | 0.67% |
| 40.9 | 3.0526(3) | 3.1033(5) | 0.80% | 0.96% |
| 44.0 | 3.0447(3) | 3.0972(5) | 0.74% | 0.82% |
| 46.2 | 3.0386(3) | 3.0934(5) | 0.83% | 1.06% |
| 51.4 | 3.0251(4) | 3.0807(6) | 0.89% | 1.18% |
| 54.0 | 3.0204(4) | 3.0784(6) | 0.87% | 1.17% |
| 56.2 | 3.0139(3) | 3.0731(5) | 0.77% | 0.93% |
| 58.0 | 3.0084(3) | 3.0712(6) | 0.75% | 0.93% |
| 60.6 | 3.0012(4) | 3.0664(7) | 0.71% | 0.86% |